\documentclass{article}
\usepackage{spconf,amsmath,graphicx,booktabs,array,placeins}
\usepackage{verbatim,url}
\graphicspath{{figures/}}

\newcommand{\Fone}{\mathrm{F1}}
\newcommand{\WER}{\mathrm{WER}}
\newcommand{\NMSE}{\mathrm{NMSE}}
\title{Interpreting and Evaluating Dynamic-Rate Speech Codec Boundaries}

\name{Han Wang\textsuperscript{*,1,3}, Jiaqi Li\textsuperscript{*,1}, Yingda Shen\textsuperscript{1}, Yuxiang Wang\textsuperscript{1}, Zhizheng Wu\textsuperscript{1,2}\thanks{* Equal contribution.}}
\address{\textsuperscript{1}The Chinese University of Hong Kong, Shenzhen\quad\textsuperscript{2}Amphion Technology Co., Ltd.\\
\textsuperscript{3}Zhejiang University}

\begin{document}
\ninept
\maketitle

\begin{abstract}
Dynamic-frame-rate neural speech codecs replace a uniform frame grid with variable-duration
tokens, making boundary placement part of the representation itself. Yet it is
unclear \textbf{what these boundaries encode} and \textbf{whether interpretable boundaries are
also useful for neural speech reconstruction}. 
This work combines boundary interpretation and controlled reconstruction
analysis by comparing predicted boundaries with linguistic and acoustic references.
We find that boundary meaning depends on the underlying speech representation. For the ASR-oriented
SenseVoice and Whisper encoders, shallow layers emphasize
phonetic, voicing, and acoustic transitions, whereas deeper layers
shift toward syllable and subword structure. In a comparison of
six dynamic-frame-rate algorithms and a uniform (fixed-frame-rate) baseline, higher-level linguistic
alignment is associated with lower pooling distortion and better reconstruction
from semantic tokens.
Frame-rate-matched boundary tests make the distinction concrete: a
syllable-derived partition improves over both Uniform and Similarity, while a
phoneme-derived partition does not improve over Uniform. We infer that for semantically rich speech representations, \textbf{useful codec boundaries are best
understood as allocation decisions organized around the syllable scale}.
\end{abstract}


\section{Introduction}
\label{sec:intro}

Neural audio codecs compress waveforms into discrete tokens; reconstruction is
the speech waveform recovered by decoding those tokens. Systems such as
SoundStream~\cite{zeghidour2021soundstream}, EnCodec~\cite{defossez2022encodec}, and DAC~\cite{kumar2023dac} improve fidelity at a given bitrate by
learning stronger encoders, quantizers, and waveform decoders, but their latent
frames remain uniformly spaced in time. This uniformity is a modeling
convenience rather than a property of speech: a steady vowel, a silence
interval, and a rapid
consonant--vowel transition need not require the same number of tokens.
Dynamic-frame-rate (DFR) codecs such as FlexiCodec~\cite{li2025flexicodec},
CodecSlime~\cite{wang2025codecslime}, and
ElasticTime~\cite{bralios2026elastictime} address this mismatch by merging
redundant frames and preserving more temporal resolution where the signal
changes.

Once the frame grid becomes adaptive, however, the representation is no longer
specified by token count alone. The codec must determine not only how many
tokens to use, but also where each token starts and ends. Existing systems
expose this design space: FlexiCodec~\cite{li2025flexicodec} uses adjacent
similarity on  semantically rich audio representations, CodecSlime~\cite{wang2025codecslime} uses dynamic programming,
ElasticTime~\cite{bralios2026elastictime} uses learned temporal bottlenecks,
DTM-Codec~\cite{sohn2026dtmcodec} uses Path Length Equalization (PLE), and
VARSTok~\cite{zheng2026varstok} uses temporal-aware density-peak clustering
(TADPC). Temporally Flexible
Coding (TFC)~\cite{zhang2025tfc} allocates temporal resolution according to
temporal entropy. Different mechanisms
can produce similar average frame rates while assigning tokens to markedly
different regions of the same utterance. These choices matter because each
boundary determines which neighboring frames are pooled
before quantization and which local changes remain as separate inputs to the
decoder.

\textit{A natural hypothesis is that useful codec boundaries align with pronunciation
units.} Phonemes and syllables are especially plausible candidates: phonemes mark
fine articulatory contrasts, whereas syllables have a coarser temporal span.
Sentence-level self-distillation in SD-HuBERT~\cite{cho2024sdhubert} induces
syllabic organization without transcript supervision.
Recent speech tokenizers such as SyllableLM~\cite{baade2025syllablelm} and
Sylber~\cite{cho2025sylber} learn syllable-like units for
efficient language modeling. It remains unclear whether stronger linguistic
alignment predicts better reconstruction when the codec interface and average
frame rate are held fixed. 
Conversely, a single labeled unit can contain multiple acoustic changes; at a
low token rate, merging the whole span into one vector may discard information.
Thus, linguistic agreement should be tested rather than assumed to imply better
reconstruction.

This work asks: \emph{what do dynamic-rate codec boundaries mean, and which
aspects of that meaning make them useful?}
For our investigation, we adopt FlexiCodec as the common codec architecture,
vary the boundary-allocation method, and fine-tune a shared decoder to
accommodate the resulting allocations.
We organize the study into two
experiments. \textbf{EXP1} evaluates boundary interpretability by measuring
agreement with phoneme, syllable, byte-pair encoding
(BPE)~\cite{sennrich2016bpe}, word, acoustic-event, and
voiced/unvoiced (V/UV) reference boundaries, then tracing the same criterion through
SenseVoice~\cite{an2024funaudiollm}, Whisper~\cite{radford2022whisper},
HuBERT~\cite{hsu2021hubert}, and WavLM~\cite{chen2022wavlm} across
representation depth. \textbf{EXP2} evaluates boundary reconstruction utility
and correlation: we compare seven methods under a fixed codec interface, relate
alignment to reconstruction quality, and test phoneme- and syllable-derived
boundaries at matched frame rates.

The results show that not all linguistic structure is equally useful for
reconstruction. Dynamic boundaries show their largest method-dependent
differences on syllable, BPE, and word boundaries, and stronger alignment with
these units tends to accompany better reconstruction. Syllable-derived
segmentation improves over both Uniform and Similarity in the matched-rate
test, whereas direct phoneme segmentation performs worse than Uniform.
Syllables thus provide a
useful organizational scale, while effective compression still requires token
placement to adapt to uneven information density within and across syllables.

\section{EXP1: Boundary Interpretability}
\label{sec:exp1}

\label{sec:meaning}

\begin{figure}[t]
  \centering
  \includegraphics[width=\columnwidth]{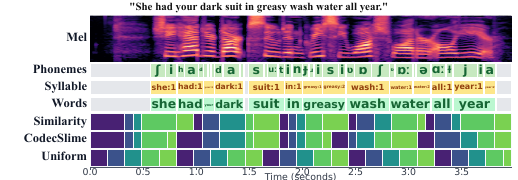}
  \caption{EXP1:\hspace{0.25em}Visualization of reference boundaries and
  rate-matched segmentations for one utterance.}
  \label{fig:boundary-vocabulary}
\end{figure}

The goal of EXP1 is to evaluate boundary interpretability against phoneme,
syllable, BPE, word, acoustic-event, and voiced/unvoiced (V/UV) reference
boundaries.
We compare six DFR algorithms and trace the criterion through encoder depth;
Fig.~\ref{fig:boundary-vocabulary} shows three reference tracks and
representative segmentations derived in EXP1.

\subsection{Data and reference-boundary construction}
\label{sec:protocol}

We use the 1,680-utterance test split of the
TIMIT dataset~\cite{garofolo1993timit} for evaluation. For each utterance, we
derive six types of reference boundaries using forced alignment, linguistic
rules, tokenization, and acoustic heuristics.
Phoneme and word intervals are
obtained with Montreal Forced Aligner (MFA) 3.0~\cite{mcauliffe2026mfa3}
and cross-checked against native TIMIT
annotations.\footnote{Using the same 40-ms matching rule, phoneme and word F1
between MFA and native TIMIT boundaries are 0.841 and 0.831, respectively. We
use MFA 3.0 rather than the TIMIT annotations because MFA 3.0 boundaries have
been shown to be more accurate~\cite{mcauliffe2026mfa3}.}
We derive syllables within each word using the Natural Language Toolkit
(NLTK)~\cite{bird2009nltk}, with vowels as nuclei.
For BPE, we tokenize each normalized word with GPT-2
and linearly map token character offsets to its MFA interval without snapping
to phoneme boundaries. 

The V/UV boundaries use 25-ms windows with a 10-ms hop on 16-kHz audio. A frame
is labeled voiced when its RMS energy exceeds an utterance-adaptive threshold
and its zero-crossing rate falls below a second utterance-adaptive threshold;
otherwise it is unvoiced.
The acoustic-event track combines changes in energy, zero-crossing rate,
spectral centroid, and spectral flatness with V/UV transitions. Further
construction details are available in the released code.\footnote{\scriptsize
Code and implementation details are available at:
\url{github.com/Hanzz3/dfr-codec-boundaries}}

Boundary F1 measures whether predicted and reference split times coincide. We
maximize the number $M$ of monotonic one-to-one matches within 40 ms, so each
boundary is counted at most once. After pooling counts across utterances, we
compute $\Fone=2M/(N_{\mathrm{pred}}+N_{\mathrm{ref}})$, where
$N_{\mathrm{pred}}$ and $N_{\mathrm{ref}}$ count predicted and reference boundaries.

\subsection{DFR algorithm configuration and rate control}

FlexiCodec~\cite{li2025flexicodec} represents speech on a 12.5-Hz latent grid
before merging frames to reach lower rates. We use this native grid so all
methods share codec inputs, quantizer, and decoder. At matched target rates, we
compare six DFR algorithms with Uniform, a fixed-rate baseline of evenly spaced
spans. Five methods operate on frozen
SenseVoice~\cite{an2024funaudiollm} layer-49 features mapped to this grid:
Similarity~\cite{li2025flexicodec}, CodecSlime~\cite{wang2025codecslime},
PLE~\cite{sohn2026dtmcodec}, TADPC from VARSTok~\cite{zheng2026varstok}, and
A-ToMe~\cite{li2023atome}. 
We also include Entropy-Mass, a baseline inspired by TFC's temporal-entropy
criterion~\cite{zhang2025tfc}; it computes frame-level waveform entropy and
uses exact-budget dynamic programming to form approximately equal-entropy spans.

Rate control uses segment budgets,
fixed merge ratios, or thresholds calibrated on development speech, yielding
6.17--6.28 Hz on TIMIT TEST at the nominal 6.25-Hz setting.
All boundary-allocation methods are training-free in this comparison.

\subsection{Boundary alignment across methods}

Table~\ref{tab:alignment-methods} compares the seven methods at 6.25 Hz.
Syllable F1 (0.380--0.557) and BPE F1 (0.360--0.512) vary substantially
across methods, whereas phoneme F1 remains within a narrow range
(0.497--0.532). Similarity and
CodecSlime lead on syllable, BPE, and word alignment; PLE has the highest F1
for phoneme, acoustic-event, and V/UV boundaries.

\begin{table*}[t]
\centering
\caption{EXP1: Boundary F1 at nominal 6.25 Hz on TIMIT TEST with a 40-ms matching
tolerance. Each cell reports F1 and its difference from Uniform,
$\mathrm{F1}/\Delta_U$, rounded to two decimals. Boldface highlights the
higher-level linguistic-track deltas for Similarity, CodecSlime, PLE, and TADPC.}
\label{tab:alignment-methods}
\setlength{\tabcolsep}{3.2pt}
\resizebox{0.82\textwidth}{!}{%
\begin{tabular}{lccccccc}
\toprule
Boundary type & Uniform & Similarity & CodecSlime & PLE & TADPC & Entropy-Mass & A-ToMe \\
\midrule
Syllable & 0.38 / 0.00 & 0.56 / \textbf{+0.18} & 0.53 / \textbf{+0.16} & 0.48 / \textbf{+0.10} & 0.49 / \textbf{+0.11} & 0.38 / +0.00 & 0.40 / +0.02 \\
BPE & 0.36 / 0.00 & 0.51 / \textbf{+0.15} & 0.48 / \textbf{+0.12} & 0.45 / \textbf{+0.09} & 0.45 / \textbf{+0.09} & 0.36 / +0.00 & 0.38 / +0.02 \\
Word & 0.29 / 0.00 & 0.45 / \textbf{+0.16} & 0.41 / \textbf{+0.12} & 0.40 / \textbf{+0.11} & 0.38 / \textbf{+0.09} & 0.29 / +0.00 & 0.30 / +0.01 \\
Phoneme & 0.50 / 0.00 & 0.52 / +0.03 & 0.52 / +0.02 & 0.53 / +0.03 & 0.50 / 0.00 & 0.50 / +0.00 & 0.50 / 0.00 \\
Acoustic event & 0.47 / 0.00 & 0.48 / +0.01 & 0.49 / +0.01 & 0.50 / +0.02 & 0.46 / -0.02 & 0.48 / +0.01 & 0.48 / +0.01 \\
V/UV & 0.44 / 0.00 & 0.43 / -0.01 & 0.43 / -0.01 & 0.46 / +0.01 & 0.41 / -0.03 & 0.45 / +0.01 & 0.44 / 0.00 \\
\bottomrule
\end{tabular}%
}
\vspace{-14pt}
\end{table*}

Absolute F1 alone is not enough to identify this preference because the six
boundary types have different densities. Phoneme boundaries are much
denser than syllable boundaries, so increasing the codec rate creates more
opportunities to match phonemes even if the DFR algorithm does not change what it
favors. To account for this counting effect, we compare against Uniform at
the same nominal target frame rate and report
$\Delta_U=\mathrm{F1}_{\mathrm{algorithm}}-\mathrm{F1}_{\mathrm{Uniform}}$
alongside every absolute value in Table~\ref{tab:alignment-methods}.

Similarity, CodecSlime, PLE, and TADPC show substantially larger
$\Delta_U$ for syllable, BPE, and word boundaries than for phonetic and
acoustic boundaries (Table~\ref{tab:alignment-methods}). A-ToMe remains
close to Uniform at 6.25 Hz because 50\% adjacent pairing often produces
two-frame spans; at 8.33 Hz, however, its syllable gain (+0.125) again
exceeds its phoneme gain (+0.029).

The same contrast persists across frame rates. Averaged over the six DFR
algorithms, mean $\Delta_U$ ranges from +0.063 to +0.100 for the three
higher-level tracks, compared with +0.002 to +0.022 for phoneme,
acoustic-event, and V/UV. For the SenseVoice representation, \textbf{most DFR algorithms therefore
preferentially place boundaries at higher-level linguistic units, including
syllables, BPE subwords, and words.}

We also compare final-layer features from different encoders on a common 12.5-Hz grid at a nominal target rate of 6.25 Hz.
Using the same Similarity rule, the ASR-oriented
SenseVoice and Whisper representations yield larger gains over Uniform for
syllable, BPE, and word boundaries than for phonetic and low-level acoustic
boundaries. For the self-supervised HuBERT and WavLM representations, the
pattern is reversed: gains are larger for phoneme, acoustic-event, and V/UV
boundaries. \textbf{Boundary selection preferences therefore depend on the encoder
representation.}

A likely explanation is that \textbf{many DFR criteria inherit the temporal
structure encoded in the underlying representation.} When ASR-oriented
representations are smoother within higher-level units and change more strongly
near their transitions, these methods tend to favor higher-level linguistic boundaries like syllables. Representations that preserve stronger local phonetic
and acoustic variation instead lead these methods to favor lower-level boundaries.

\subsection{Representation depth changes boundary meaning}
\label{sec:depth}

\begin{figure}[t]
  \centering
  \includegraphics[width=\columnwidth]{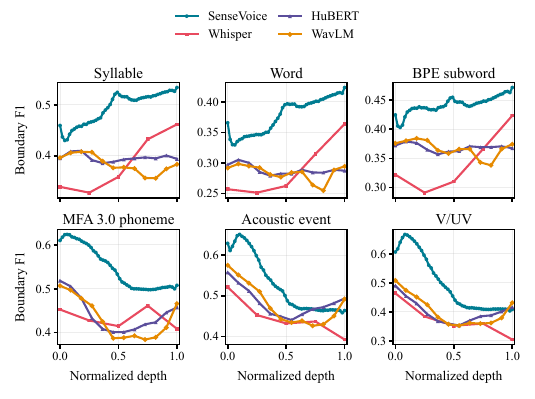}
  \caption{EXP1: Boundary F1 across normalized encoder depth at 6.25 Hz and a 40-ms
  tolerance window. Panels use independent y-axis scales.}
  \label{fig:layers}
\end{figure}

We keep the adjacent-similarity criterion and target rate of 6.25 Hz fixed
while varying encoder depth. Fig.~\ref{fig:layers} reports alignment on each
encoder's native time grid for SenseVoice~\cite{an2024funaudiollm},
Whisper~\cite{radford2022whisper}, HuBERT~\cite{hsu2021hubert}, and
WavLM~\cite{chen2022wavlm}. In SenseVoice, syllable, BPE, and word F1
generally increase toward deeper layers, whereas phoneme, acoustic-event,
and V/UV F1 peak in shallow layers. Whisper also reaches its highest
syllable, BPE, and word F1 in the final layer, while acoustic-event and V/UV
alignment are lower than in the shallow layers.

HuBERT and WavLM show a different pattern. Their phoneme, acoustic-event,
and V/UV F1 decline through the early and middle layers, then partly recover
near the output, remaining below their shallow-layer values. Syllable, BPE,
and word F1 fluctuate without a comparable upward trend.

These depth-wise trends are consistent with the representation-dependent explanation above. In SenseVoice and Whisper, transcription supervision increasingly emphasizes higher-level linguistic structure in deeper layers, leading to stronger syllable, BPE, and word alignment. HuBERT and WavLM, by contrast, do not show the same linguistic shift: local phonetic and acoustic alignment weakens with depth, while higher-level alignment does not increase consistently. \textbf{Greater depth therefore changes which temporal structures are exposed to boundary selection: it increasingly emphasizes higher-level linguistic organization in ASR-oriented encoders, while primarily suppressing local acoustic structure in self-supervised encoders.}

\section{EXP2: Boundary Reconstruction Evaluation and Correlation}
\label{sec:utility}

We next examine whether the boundary structure identified above is useful for
reconstruction. Under a shared codec interface, we compare seven methods,
examine cross-rate associations between boundary F1 and reconstruction utility,
and compare phoneme- and syllable-derived segmentations with rate-matched
controls.

\subsection{Reconstruction across boundary-allocation methods}
\label{sec:controlled-recon}

{\looseness=-1
We compare the seven methods at the nominal 6.25-Hz operating point,
keeping SenseVoice~\cite{an2024funaudiollm} layer-49 features and the
FlexiCodec~\cite{li2025flexicodec} encoder fixed. All methods share one
waveform decoder, fine-tuned on LibriTTS train-clean-100 with a balanced
mixture of boundary allocations at 3, 6.25, 8.33, and 10 Hz, then held fixed
for evaluation. We reconstruct
speech using the first through eighth residual vector quantization (RVQ)
codebooks (q1:8) or only the first
codebook (q1).\par}

{\looseness=-1
For each method, q1 and q1:8 use the same selected spans. We evaluate q1 and q1:8 reconstruction WER with Whisper Small~\cite{radford2022whisper}, and q1 ASR WER using a Qwen2.5~\cite{qwen2024qwen25} transcriber trained directly on q1 tokens. For q1:8 reconstruction, we also report PESQ and speaker similarity using WavLM-Base-Plus~\cite{chen2022wavlm} embeddings. NMSE measures pre-quantization pooling distortion between the original features and their segment-wise means. Runtime includes method-specific scoring and boundary selection.\par}

\begin{figure}[!b]
  \centering
  \includegraphics[width=0.84\columnwidth]{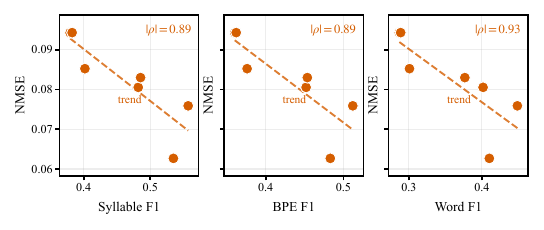}
  \setlength{\abovecaptionskip}{0pt}
  \caption{EXP2: Syllable, BPE, and word alignment versus feature NMSE across
  seven methods at 6.25 Hz. Dashed lines are linear-fit guides; each panel
  reports absolute Spearman correlation.}
  \label{fig:alignutility}
\end{figure}

\begin{table*}[t]
\centering
\caption{EXP2: Reconstruction quality and boundary-selection runtime for seven
methods at 6.25 Hz on TIMIT TEST. WER is corpus-level; runtime excludes input
loading and encoder extraction.}
\label{tab:recon}
\setlength{\tabcolsep}{4.2pt}
\resizebox{0.94\textwidth}{!}{%
\begin{tabular}{lcccccccc}
\toprule
Method & Syllable F1/Rank & q1 ASR WER$\downarrow$ & q1 recon WER$\downarrow$ & q1:8 recon WER$\downarrow$ & NMSE$\downarrow$ & PESQ$\uparrow$ & SpkSim$\uparrow$ & Runtime (ms)$\downarrow$ \\
\midrule
Similarity & 0.557 / 1 & \underline{7.762} & 13.304 & 5.979 & \underline{0.0759} & 2.294 & \underline{0.9247} & \underline{0.277} \\
CodecSlime & 0.535 / 2 & \textbf{7.586} & \textbf{10.411} & \textbf{5.394} & \textbf{0.0627} & 2.343 & 0.9246 & 19.546 \\
TADPC & 0.485 / 3 & 8.438 & 16.328 & 5.731 & 0.0830 & 2.316 & \textbf{0.9248} & 1.945 \\
PLE & 0.482 / 4 & 8.397 & \underline{12.967} & \underline{5.415} & 0.0805 & 2.330 & 0.9238 & 0.283 \\
A-ToMe & 0.402 / 5 & 8.269 & 16.953 & 6.116 & 0.0852 & \textbf{2.360} & 0.9225 & 1.290 \\
Entropy-Mass & 0.383 / 6 & 8.417 & 20.293 & 6.968 & 0.0943 & \underline{2.349} & 0.9207 & 31.017 \\
Uniform & 0.380 / 7 & 8.168 & 20.004 & 6.226 & 0.0943 & 2.340 & 0.9211 & \textbf{0.016} \\
\bottomrule
\end{tabular}
}
\end{table*}

{\looseness=-2
Table~\ref{tab:recon} shows that boundary placement substantially affects reconstruction at comparable frame rates. CodecSlime achieves the lowest WERs and NMSE but incurs a high boundary-selection cost (19.546 ms) because it scores candidate spans and performs dynamic-programming search. PLE reaches nearly the same q1:8 reconstruction WER (5.415\% vs. 5.394\%) at a much lower cost (0.283 ms). A-ToMe and TADPC achieve the best PESQ and SpkSim, respectively. Despite the larger degradation when using q1 alone, q1 and q1:8 method rankings remain strongly correlated ($\rho=0.964$).\par}

\subsection{Higher-level linguistic alignment and reconstruction}
\label{sec:correlation}

{\looseness=-2
Table~\ref{tab:alignment-methods} shows that syllable, BPE, and word F1 vary
substantially across methods, whereas phoneme, acoustic-event, and V/UV F1
occupy much narrower ranges. We therefore ask whether the higher-level
linguistic alignment that distinguishes the methods is related to reconstruction
utility. At 6.25 Hz, Fig.~\ref{fig:alignutility} shows a clear rank-level
association: methods with stronger syllable, BPE, and word alignment exhibit
lower pre-quantization pooling distortion.\par}

\begin{table}[t]
\centering
\caption{EXP2: Cross-rate absolute Spearman correlations $|\rho|$ between
syllable/BPE/word boundary F1 and reconstruction utility across seven methods.}
\label{tab:srcc}
\setlength{\tabcolsep}{3.2pt}
\resizebox{\columnwidth}{!}{%
\begin{tabular}{lccc}
\toprule
Rate & q1 recon $\WER$  & $\NMSE$ & SpkSim \\
\midrule
6.25 Hz & 0.79/0.79/0.86 & 0.89/0.89/0.93 & 0.86/0.86/0.75 \\
8.33 Hz & 0.89/0.82/0.82 & 0.71/0.61/0.61 & 0.54/0.57/0.57 \\
10 Hz   & 0.82/0.82/0.79 & 0.79/0.79/0.75 & 0.64/0.64/0.61 \\
\bottomrule
\end{tabular}%
}
\end{table}

Table~\ref{tab:srcc} extends this relationship across 6.25, 8.33, and 10 Hz.
For all three higher-level boundary types, alignment remains consistently
associated with lower reconstruction WER and lower NMSE across the tested
rates. The association with SpkSim is weaker at higher rates, where speaker
similarity is already near saturation and varies little across methods. At
6.25 Hz, the same NMSE trend is also observed on LibriSpeech test-clean, with
identical method rankings.

\textbf{Taken together, these results show that higher-level linguistic
alignment is consistently associated with reconstruction utility across
methods and frame rates.} Stronger syllable alignment, in particular,
accompanies lower reconstruction WER and lower pooling distortion. This motivates a direct test of whether linguistic boundaries themselves form useful codec segments and whether the unit scale matters.

\subsection{Frame-rate-matched linguistic boundary tests}
\label{sec:linguistic-boundary-tests}

To test this directly, we merge frames within each pronunciation unit into one
token. Phonemes and syllables provide two natural
partitions, which we compare with rate-matched controls. The phoneme-derived partition operates at approximately
8.06 Hz after projection onto the codec grid, where nearby boundaries can collapse
onto the same frame. The syllable-derived partition operates at approximately
4.47 Hz. We use Uniform as a fixed-rate control and CodecSlime and Similarity
as adaptive controls at the corresponding nominal rates.

Table~\ref{tab:linguistic-boundary-tests} shows opposite effects relative to
Uniform. At 8.06 Hz, phoneme-derived boundaries increase WER from 6.47\% to
7.10\%, and both adaptive controls perform better. At 4.47 Hz,
syllable-derived boundaries reduce WER from 17.61\% to 13.85\% and outperform
Similarity (14.99\%), although CodecSlime remains better (11.28\%).
Under this shared SenseVoice representation, \textbf{syllable-derived
boundaries improve reconstruction, whereas phoneme-derived boundaries do not.}

\begin{table}[!t]
\centering
\caption{EXP2: Corpus WER (\%) for reference-derived and rate-matched
segmentations on TIMIT TEST at phoneme (8.06 Hz) and syllable (4.47 Hz) rates.
Bold/underline mark the best/second-best result within each rate.}
\label{tab:linguistic-boundary-tests}
\setlength{\tabcolsep}{4.8pt}
\begin{tabular}{lrr}
\toprule
Segmentation & Phoneme-rate WER$\downarrow$ & Syllable-rate WER$\downarrow$ \\
\midrule
Uniform           & 6.47 & 17.61 \\
Reference-derived & 7.10 & \underline{13.85} \\
CodecSlime        & \underline{5.11} & \textbf{11.28} \\
Similarity        & \textbf{5.05} & 14.99 \\
\bottomrule
\end{tabular}
\end{table}

One possible explanation is that phoneme transitions do not always coincide
with changes that warrant separate codec tokens: coarticulation can extend
across a boundary, while substantial variation can occur within a phoneme.
Syllable-scale grouping may better preserve such local continuity. However,
syllables also differ in duration and internal complexity, and a fixed syllable
partition cannot allocate additional tokens within a complex syllable. These
results refine Sec.~\ref{sec:correlation}: linguistic alignment alone is
insufficient, and the unit scale also matters. Syllable-scale grouping aids
reconstruction, while the remaining gap to CodecSlime shows that adaptive
placement can still improve on fixed syllable boundaries.

\FloatBarrier
\section{Conclusion}
\label{sec:conclusion}
We evaluated dynamic-rate speech codec boundaries using six types of reference
boundaries and controlled reconstruction tests. Boundary meaning depends on the
encoder representation and its depth: deeper ASR-oriented encoders increasingly
align with higher-level linguistic structure, whereas self-supervised encoders
show declining acoustic alignment without a comparable shift toward linguistic
units. Across three rates, stronger syllable alignment accompanies better
reconstruction, as reflected by lower reconstruction WER. It is also
associated with lower pooling distortion.
Among the two fixed pronunciation partitions tested, syllable-derived
segmentation improves over rate-matched Uniform, while phoneme-derived
segmentation does not. Together, these results suggest treating
syllable-scale organization as a useful prior rather than a fixed tokenization
rule: it can guide token allocation, while adaptive placement captures variation
within and across syllables. 

However, our analysis covers only English speech,
four encoders, seven methods, and three target frame rates. In particular, the
12.5-Hz output grid of SenseVoice limits our evaluation of higher-rate settings,
so the observed relationships are not established above this rate. 
These conclusions are also limited to semantically rich encoder features because
the acoustic representations considered here are incompatible with the current
FlexiCodec framework. Future work will extend the analysis to higher frame
rates, broader representation families, and multilingual speech.

\section{Acknowledgment}
This work is partially supported by NSFC Grant 62376237; 2023 Shenzhen Stability Science Program; Program for Guangdong Introducing Innovative and Entrepreneurial Teams 2023ZT10X044.

OpenAI ChatGPT was used for language and grammar checking, figure and table
preparation, and assistance with experimental code. All research ideas and
scientific conclusions are the authors' own; they verified and take full
responsibility for the manuscript and accompanying artifacts.

\section{Compliance with Ethical Standards}
This study used the existing TIMIT, LibriSpeech and LibriTTS speech datasets under their applicable licenses
and involved no new data collection or interaction with human participants. No
additional ethical approval was required.

\FloatBarrier

\bibliographystyle{IEEEbib}
\bibliography{boundary_icassp_refs}

\end{document}